# Local polynomial solutions of steady Euler equations for planar ideal fluids

Wennan Zou [*1]

[1] Institute of Fluid Mechanics, Nanchang University, Nanchang 330031, China
*Email: zouwnn@ncu.edu.cn

**ABSTRACT**：Exploring the general analytical solutions to the Euler equations for ideal fluids holds significant theoretical and practical importance. The steady flows in two-dimensional spaces are considered whether there is an analytical solution in the form of a finite polynomial within the local region. By employing the tensorial representation and the complex variable notation, we successfully work out the local analytical solutions in terms of polynomials with highest degree up to four. Several examples are illustrated and some discussion comments on the effect of the viscous term from the Navier-Stokes (N-S) equations are presented.

**Key Words**：Euler equations; local analytical solutions; convective term; tensorial representation; complex variable method

## 1 Introduction

The Euler equations, governing inviscid incompressible flows since their inception in 1761 (Euler, 1761), remain a cornerstone of fluid dynamics. Despite their apparent simplicity, the nonlinear convective term poses profound mathematical challenges, restricting analytical solutions primarily to flows with decoupled nonlinearity—irrotational velocity fields, two-dimensional (2D) constant vorticity domains, or three-dimensional (3D) Beltrami flows, etc. (Majda and Bertozzi, 2002). But this is not enough! Reynolds (1895) proposed the concept of quadratic averaging and obtained the "turbulent stress" from the nonlinear convective term. In all textbooks, the derived vorticity equation is interpreted by dividing the convective term into two parts, where the stretching part is speculated to cause finite time singularities (Pumir and Siggia, 1992; Hou and Li, 2008). Upon to now, the core of the nonlinear convective term remains largely a mystery, and a hurdle that all flow theories cannot overcome.

The recent studies on the Euler equations are mainly led by mathematicians, focusing on discussing finite time singularities. The theoretical advance extended from 2D to 3D, and first formed some breakthrough on 3D axisymmetric flows through numerical method (Majda and Bertozzi, 2002; Gibbon, 2008; Luo and Hou, 2014). Along with the development of numerical technology, the simulation of Euler equations increasingly proves its power (Hou and Li, 2008; Grafke *et al*., 2008; Gibbon, 2008; Luo and Hou, 2019). It is believed that future progress on the understanding of Euler equations will also rely on the interplay of theory and simulation. On the other hand, it is well known that closed-form solutions retain indispensable value: they provide benchmark tests, reveal intrinsic flow topologies, and elucidate mechanisms of energy transfer and vortex formation; but relatively speaking the exact solutions with finite form are less reported, especially for 3D Euler equations. Bajer and Moffatt (1990) studied a 3D quadratic velocity field satisfying the incompressibility condition and bounded within a sphere. Following Brøns (1994) and Hartnack (1999), Deliceoğlu and Gürcan (2008) tried to make a 2D Hamiltonian system in series form and with mirror symmetry about the *y*-axis to satisfy the Navier-Stokes (N-S) equations, and classified the local streamline topologies of velocity fields with certain physical authenticity. However, they all circumvent the core analytical challenge: resolving nonlinearity algebraically in finite-dimensional function spaces.

The motivation of this study is to work out exact local solutions with finite polynomial form for the steady Euler equations by deconstructing the nonlinear convective term. Unlike the proposition summarized by Majda and Bertozzi (2002), which is especially suitable for 2D Euler equations, we will present a systematic methodology for deriving finite-degree polynomial solutions that is applicable to both the 2D and 3D steady Euler equations in local domains. Several mathematical means, such as tensor representation and decomposition, complex variable method and homotopy operator, are combined and applied throughout the process, and the physical constraints from the Euler equations are strictly satisfied. As the first implementation of such a solution strategy, this paper will focus on the 2D steady Euler equations.

In the remainder of this paper, we first present the compact form of the Euler equations and mathematical means to be used. The velocity field in terms of finite polynomials is reduced to an expression with less independent coefficients by satisfying the incompressible condition, and the vorticity is expressible with partial coefficients of the velocity. In virtue of the homotopy operation and decomposition in a starshaped region, the total pressure can be explicitly expressed by the coefficients of the velocity, and the irrotational property of the convective term results in a set of nonlinear coupling relations between the coefficients, which can be solved as explicit constraints. The finite solutions of velocity field finally take three types: all velocity coefficients being nonzero and strongly coupled, the vorticity being radially distributed and the velocity being harmonic polynomials. A brief conclusion is given after some discussions are presented.

## 2 Euler equations and mathematical preparation

### 2.1 Euler equations

In this paper, the Cartesian coordinate system $\{O; x_i, i = 1,2\}$ is adopted in the two-dimensional spaces, where the origin is set to be the center of a local open region occupied by the ideal fluid. The shape of the region is assumed to be starshaped with respect to some points belong to the domain, including the origin. In addition, uses are made of the Einstein's summation convention for the indices associated with the base vectors $\{\mathbf{e}_i, i = 1,2\}$ of coordinate system, $\delta_{ij} = \{1, \text{if } i = j; 0, \text{else}\}$ and $\epsilon_{ij} = \{0, \text{for } i = j; 1, \text{for } (i,j) = (1,2); -1, \text{for } (i,j) = (2,1)\}$ are used to indicate the Kronecker delta and the permutation symbol, respectively. Thus, by taking the value of the density to be one, the continuity equation and the steady Euler equations read (Majda and Bertozzi, 2002)

$$\begin{cases} \partial_k v_k = 0, \\ v_k \partial_k v_i = -\partial_i p. \end{cases} \tag{1}$$

Here $v_i, p$ indicate the velocity field and the pressure field, respectively. Any body force with a potential, say $f_i = -\partial_k \phi$, can be contained in the pressure field.

In planar flows, since the direction of vorticity is fixed perpendicular to the plane, so it can be regarded as a scalar, and expressed by

$$\Omega = \epsilon_{ij} \partial_i v_j. \tag{2}$$

Making use of the vorticity and introducing the total pressure

$$P = p + \frac{1}{2} v_i v_i, \tag{3}$$

the Euler equations $(1)_2$ can be written in an equivalent form, namely

$$\Omega v_k \epsilon_{ki} = -\partial_i P. \tag{4}$$

In our subsequent derivation, the latter expression form (4) is more convenient to use.

### 2.2 Differential form and its homotopy decomposition

Different form is a fundamental mathematical language in the studies of geometry and topology. In order to better utilize modern mathematical tools, we can flexibly express physical fields and their equations with differential forms instead of vectors. For example, the velocity can be treated as a differential 1-form field by

$$\tilde{v} = \boldsymbol{v} \cdot d\boldsymbol{x} = v_k dx_i, \tag{5}$$

while the vorticity can be written by a differential 2-form field as the exterior differential of $\tilde{v}$, say

$$\tilde{\Omega} = d\tilde{v} = \partial_i v_k dx_i \wedge dx_k = \partial_i v_k \epsilon_{ik} da = \Omega da, \tag{6}$$

where the basic differential 2-form $da = dx_1 \wedge dx_2$ indicates the area element. Starting from the velocity 1-form in three-dimensional spaces, the set $\{I_1 = \tilde{v}, I_2 = dI_1 = \tilde{\Omega}, I_3 = \tilde{v} \wedge I_2 = \tilde{h}\}$ constitute a chain, the integer $m$ from $I_m \neq 0, I_{m+1} = 0$ can be used to characterize the function dimension of the velocity, which could be further defined locally or globally. Back to 2D cases, the maximal function dimension of velocity field is 2.

All differential forms in 2D spaces $E_2$ constitute the space $\Lambda(E_2) = \Lambda^0(E_2) \oplus \Lambda^1(E_2) \oplus \Lambda^2(E_2)$. A differential form $\alpha$ is said to be *closed* iff $d\alpha = 0$, and to be *exact* iff $\alpha = d\beta$. The inner multiplication $\boldsymbol{a} \rfloor \alpha$ between a vector $\boldsymbol{a}$ and a differential form $\alpha$, is defined from their bases: $\mathbf{e}_i \rfloor dx_k = \delta_{ij}$, where the differential form can be generalized to any order, say $\mathbf{e}_i \rfloor 1 = 0, \mathbf{e}_i \rfloor dx_j \wedge dx_k = \langle dx_j \wedge dx_k, \mathbf{e}_i \rangle = \delta_{ij} dx_k - \delta_{ik} dx_j$. We study differential forms in a starshaped region $\mathcal{D}$ with the point $x_i^0$ as one of its centers, that means for any point $x_k \in \mathcal{D}$ we have $x_i^0 + \lambda(x_k - x_i^0) \in \mathcal{D}$ for all $\lambda \in [0,1]$. Thus, from $\alpha(x_i) = \alpha_{i_1 \cdots i_p}(x_i) dx_{i_1} \wedge \cdots \wedge dx_{i_p}$, we have $\hat{\alpha}(\lambda) = \alpha_{i_1 \cdots i_p}\left(x_i^0 + \lambda(x_i - x_i^0)\right) dx_{i_1} \wedge \cdots \wedge dx_{i_p}$ associated with the local radius vector $\boldsymbol{X}(\boldsymbol{x}) = \boldsymbol{x} - \boldsymbol{x}_0$. In particular, we have properties

$$\hat{\alpha}(1) = \alpha, d\hat{\alpha}(\lambda) = \lambda(\widetilde{d\alpha})(\lambda), \qquad \widehat{\boldsymbol{X}}(\lambda) = \boldsymbol{X}(\boldsymbol{x}_0 + \lambda \boldsymbol{X}) = \lambda \boldsymbol{X}. \tag{7}$$

The linear homotopy operator $\mathcal{H}$ for a $K$th-degree differential form $\alpha$ on a starshaped region $\mathcal{D}$ is defined by

$$\mathcal{H}\alpha = \int_0^1 \boldsymbol{X} \rfloor \tilde{\alpha}(\lambda) \lambda^{K-1} d\lambda = \int_0^1 \boldsymbol{X}(\boldsymbol{x}) \rfloor \alpha\left(x_{0i} + \lambda(x_i - x_i^0)\right) \lambda^{K-1} d\lambda. \tag{8}$$

The operator $\mathcal{H}$ has the following useful properties: (H1) $\mathcal{H}$ maps $\Lambda^K(\mathcal{D})$ into $\Lambda^{K-1}(\mathcal{D})$ for $K \geq 1$ and $\Lambda^0(\mathcal{D})$ into zero; (H2) $d\mathcal{H} + \mathcal{H}d =$ identity for $K \geq 1$, and $(\mathcal{H}df)(x_i) = f(x_i) - f(x_i^0)$; (H3) $(\mathcal{H}\mathcal{H}\alpha)(x_i) = 0$, $(\mathcal{H}\alpha)(x_i^0) = 0$; (H4) $\mathcal{H}d\mathcal{H} = \mathcal{H}, d\mathcal{H}d = d$; (H5) $\mathcal{H}d\mathcal{H}d = \mathcal{H}d, d\mathcal{H}d\mathcal{H} = d\mathcal{H}, (d\mathcal{H})(\mathcal{H}d) = 0, (\mathcal{H}d)(d\mathcal{H}) = 0$; (H6) $\boldsymbol{X} \rfloor \mathcal{H} = 0, \mathcal{H}\boldsymbol{X} \rfloor = 0$. After the above preparation, we may introduce the homotopy operator. It is easy to know that $d$ is the inverse of $\mathcal{H}$ when the domain of $\mathcal{H}$ is limited to the space $\mathcal{E}^K(\mathcal{D}), K \geq 1$ of all exact forms, which can be used to identify the exact part $\alpha_e$ of $\alpha \in \Lambda^K(\mathcal{D})$ to be $d\mathcal{H}\alpha$. The remainder $\alpha - \alpha_e$ belongs to the kernel of linear operator $\mathcal{H}$, called the antiexact part $\alpha_a$ of $\alpha$. All antiexact forms on $\mathcal{D}$ consist of $\mathcal{A}(\mathcal{D})$, and $\Lambda(\mathcal{D}) = \mathcal{E}(\mathcal{D}) \oplus \mathcal{A}(\mathcal{D})$, which yields the crucial homotopy decomposition of differential form $\alpha$ on $\mathcal{D}$, that is

$$\alpha = \alpha_e + \alpha_a = d\mathcal{H}\alpha + \mathcal{H}d\alpha. \tag{9}$$

Different from the exterior differential operation, the implement of homotopy operator depends on the choose of the center. Let $\mathcal{H}_1$ and $\mathcal{H}_2$ be two homotopy operators on $\mathcal{D}$ with different centers, it is easy for $\alpha \in \Lambda^K(\mathcal{D})$ to prove

$$\mathcal{H}_1\alpha = \mathcal{H}_2\alpha + \beta + d\gamma, \beta = -\mathcal{H}_1\mathcal{H}_2 d\alpha, \gamma = \begin{cases} \mathcal{H}_2\mathcal{H}_1\alpha, K > 1; \\ const., K = 1. \end{cases} \tag{10}$$

**2.3 Tensorial representation of velocity field in terms of polynomials**

The velocity fields in terms of finite polynomials with maximal degree $N$ can be expressed by

$$\boldsymbol{v} = \sum_{K=0}^{N} \boldsymbol{v}^{(K)} = \sum_{K=0}^{N} \boldsymbol{x}^{\otimes K} \circ \boldsymbol{D}^{(K+1)}, \tag{11}$$

with $\boldsymbol{v}^{(K)}$ indicating a vector-valued homogeneous polynomial of degree $K$, and $\boldsymbol{D}^{(K)}$ indicating a tensor of order $K$ that is symmetric in the first $K$ indices.

According to the tensor representation theory in two-dimensional space (Zou *et al*, 2001), the tensor $\boldsymbol{D}^{(K+1)}$ may consist of two symmetric tensors of orders $p-1$ and $p+1$, i.e.

$$\boldsymbol{D}^{(K+1)} = \boldsymbol{S}^{(K+1)} \dotplus \boldsymbol{S}^{(K-1)}, \tag{12.1}$$

or in the accurate index expression

$$D_{i_1 \cdots i_K i} = S_{i_1 \cdots i_K i} + S_{\hat{i}_1 \cdots \hat{i}_{K-1}} \epsilon_{\hat{i}_K i} \tag{12.2}$$

where the symmetrization of tensor $T_{i_1 \cdots i_K i}$ on the indices $(i_1 \cdots i_K)$ is denoted by $T_{\hat{i}_1 \cdots \hat{i}_K i}$, meaning the $K!$ sum of all the permutations of $T_{i_1 \cdots i_K i}$ on the indices $(i_1 \cdots i_K)$ divided by $K!$. For instance,

$$T_{\hat{i}_1 \hat{i}_2 i} = \frac{1}{2}(T_{i_1 i_2 i} + T_{i_2 i_1 i}), u_{\hat{i}_1} v_{\hat{i}_2} = \frac{1}{2}(u_{i_1} v_{i_2} + v_{i_1} u_{i_2}). \tag{13}$$

A $p$th-order completely symmetric tensor $\boldsymbol{S}^{(K)}$ can be further decomposed into a set of symmetric traceless tensors or deviators in the form (Zheng and Zou, 2000)

$$S^{(M)} = \sum_{I=0}^{\left[\frac{M}{2}\right]} \text{sym}\left(\mathbf{1}^{\otimes I} \otimes d^{(M-2I)}\right), \tag{14}$$

where $\text{sym}(\mathbf{T})$ indicates the completely symmetric part of the tensor $\mathbf{T}$. For examples,

$$S_{ij} = d_{ij} + \alpha \delta_{ij}, \tag{15.1}$$
$$S_{ijk} = d_{ijk} + \delta_{\hat{\imath}\hat{\jmath}} u_{\hat{k}}, \tag{15.2}$$
$$S_{ijkl} = d_{ijkl} + \delta_{\hat{\imath}\hat{\jmath}} d_{\hat{k}\hat{l}} + \alpha \delta_{\hat{\imath}\hat{\jmath}} \delta_{\hat{k}\hat{l}}, \tag{15.3}$$
$$S_{ijklm} = d_{ijklm} + \delta_{\hat{\imath}\hat{\jmath}} d_{\hat{k}\hat{l}\hat{m}} + \delta_{\hat{\imath}\hat{\jmath}} \delta_{\hat{k}\hat{l}} u_{\hat{m}}. \tag{15.4}$$

And their expressions after a contraction become

$$S_{rr} = 3\alpha, S_{irr} = \tfrac{4}{3} u_i, S_{ijrr} = d_{ij} + \tfrac{4}{3}\alpha \delta_{ij}, S_{ijkrr} = \tfrac{4}{5} d_{ijk} + \tfrac{6}{5} \delta_{\hat{\imath}\hat{\jmath}} u_{\hat{k}}. \tag{15.5}$$

In the following, for a $K$-th order symmetric tensor $\boldsymbol{S}^{(N)}$ and a vector $\boldsymbol{a}$, we define an associated symmetric tensor as

$$\boldsymbol{S}_a^{(N,K)} = \boldsymbol{a}^{\otimes K} \circ \boldsymbol{S}^{(N)}, K \leq N. \tag{16}$$

Then, the expansion (11) can be rewritten by

$$\boldsymbol{v} = \sum_{K=0}^{N} \boldsymbol{S}_x^{(K+1,K)} + S_x^{(K-1,K-1)} \boldsymbol{x} \cdot \boldsymbol{\epsilon}. \tag{17}$$

### 2.4 Complex representation of velocity field and deviators in 2D spaces

In 2D flows, a more compact representation of velocity field in terms of finite polynomials is to make use of the complex variable. From $z = x_1 + \iota x_2 = \boldsymbol{w} \cdot \boldsymbol{x}$ with $\iota = \sqrt{-1}$, $\boldsymbol{w} = \boldsymbol{e}_1 + \iota \boldsymbol{e}_2$, we have

$$\frac{\partial}{\partial z} = \frac{1}{2}\left(\frac{\partial}{\partial x_1} - \iota \frac{\partial}{\partial x_2}\right), \nabla^2 = \frac{1}{4}\frac{\partial^2}{\partial z \partial \bar{z}}. \tag{18}$$

By the definition $v = v_1 + \iota v_2 = \boldsymbol{w} \cdot \boldsymbol{v}$, it is easy to derive the complex expressions of the divergence and the curl of the velocity field

$$\partial_i v_i + \iota \Omega = 2 \frac{\partial v}{\partial z}. \tag{19}$$

Since there are only two independent components for a deviator $\boldsymbol{d}^{(K)}$ of any order $K > 0$, we can introduce a complex parameter $c_K$ to indicate it, namely

$$\text{Re}(c_K) = d_{1\cdots 111} = -d_{1\cdots 122}, \text{Im}(c_K) = d_{1\cdots 112} = -d_{1\cdots 222}. \tag{20}$$

It has no difficulty to verify

$$\boldsymbol{d}^{(K)} = \text{Re}(\bar{c}_K \boldsymbol{w}^{\otimes K}) \leftrightarrow c_K = 2^{1-K} \boldsymbol{w}^{\otimes K} \circ \boldsymbol{d}^{(K)}, \tag{21}$$

and

$$\boldsymbol{x}^{\otimes K} \circ \boldsymbol{d}^{(K)} = \text{Re}(\bar{c}_K z^K), \boldsymbol{w} \cdot \left(\boldsymbol{x}^{\otimes K} \circ \boldsymbol{d}^{(K+1)}\right) = c_{K+1} \bar{z}^K. \tag{22}$$

## 3 Solution of Euler equations in terms of finite polynomials

### 3.1 Tensor formulation and solvability

When the velocity field is assumed to be a vector-valued homogeneous polynomial of degree $K$, namely

$$v_i^K = x_{i_1} \cdots x_{i_K} D_{i_1 \cdots i_K i}^K = x_{i_1} \cdots x_{i_K} S_{i_1 \cdots i_K i}^K + x_{i_1} \cdots x_{i_{K-1}} S_{i_1 \cdots i_{K-1}}^K x_{i_K} \epsilon_{i_K i}, \tag{23}$$

the incompressible condition $(1)_1$ yields

$$\partial_i v_i^K = K x_{i_1} \cdots x_{i_{K-1}} S_{i_1 \cdots i_{K-1} ii}^K + (K-1) x_{i_1} \cdots x_{i_{K-2}} S_{i_1 \cdots i_{K-2} i}^K x_{i_{K-1}} \epsilon_{i_{K-1} i} = 0. \tag{24}$$

Since this condition should be satisfied for all positions $\boldsymbol{x} \in \mathcal{D}$, so we have

$$S_{ii}^1 = 0; \quad S_{i_1 \cdots i_{K-1} ii}^K = -\frac{K-1}{K} S_{i_1 \cdots \hat{\imath}_{K-2} i}^K \epsilon_{\hat{\imath}_{K-1} i}, K > 1. \tag{25}$$

Combining with (14), (25) means that, except for the highest one, all deviators in the $(K+1)$-order symmetric tensor can be given or determined by the $(K-1)$-order symmetric tensor. Thus, after applying the incompressible condition, the homogeneous polynomial velocity of degree $K$ can be described by a set of deviators $\{\boldsymbol{d}^{K,(K+1)}, \boldsymbol{d}^{K,(K-1)}, \boldsymbol{d}^{K,(K-3)}, \cdots\}$ with the number of independent parameters to be $2\left[\frac{K+1}{2}\right] + 1$ for $K = 2M$ or $2\left[\frac{K+1}{2}\right]$ for $K = 2M + 1$. If we regard (23) as a differential 1-form and using the homotopy operator $\mathcal{H}$ with the origin as its center, it is easy to know

$$\tilde{v}^K = v_i^K dx_i = \tilde{v}_e^K + \tilde{v}_a^K, \qquad \tilde{v}_e^K = x_{i_1}\cdots x_{i_K} S_{i_1\cdots i_K i}^K dx_i, \tilde{v}_a^K = x_{i_1}\cdots x_{i_{K-1}} S_{i_1\cdots i_{K-1}}^K x_{i_K}\epsilon_{i_K i} dx_i. \tag{27}$$

Further from

$$\tilde{\Omega}^K = d\tilde{v}^K = d\tilde{v}_a^K = (K+1)x_{i_1}\cdots x_{i_{K-1}} S_{i_1\cdots i_{K-1}}^K da, \tag{28.1}$$

we obtain

$$\Omega^K = x_{i_1}\cdots x_{i_{K-1}}\vartheta_{i_1\cdots i_{K-1}}^K, \vartheta_{i_1\cdots i_{K-1}}^K = (K+1)\epsilon_{ji} D_{i_1\cdots i_{K-1}ji}^K = (K+1)x_{i_1}\cdots x_{i_{K-1}} S_{i_1\cdots i_{K-1}}^K, K > 0 \tag{28.2}$$

having nothing to do with the first term $S_{i_1\cdots i_K i}^K$ in (23)$_2$, and the tensor $\vartheta_{i_1\cdots i_{K-1}}^K$ is symmetric in all its indices. So, the tensorial decomposition (12) of the velocity naturally separates the rational part from the potential part. To sum up, we list the tensorial coefficients $D_{i_1\cdots i_K i}^K, \vartheta_{i_1\cdots i_{K-1}}^K$ and the velocity field $v_i^K$ in terms of a set of deviators in Table 1.

Table 1. Tensorial expressions of homogeneous polynomial velocities $v_i^K$ satisfying $\partial_i v_i^K = 0$

| $K$ | $D_{i_1\cdots i_K i}^K$ | $\vartheta_{i_1\cdots i_{K-1}}^K$ | $v_i^K$ |
|---|---|---|---|
| 1 | $d_{ji}^1 + \alpha^1 \epsilon_{ji}$ | $2\alpha^1$ | $x_j d_{ji}^1 + \alpha^1 x_j \epsilon_{ji}$ |
| 2 | $d_{jki}^2 - \frac{3}{8}\delta_{ij}\epsilon_{kr}u_r^2 + u_j^2\epsilon_{ki}$ | $3u_i^2$ | $x_j x_k d_{jki}^2 - \frac{1}{8}(R^2\epsilon_{ir}u_r^2 + 2x_k\epsilon_{kr}u_r^2 x_i) + x_j u_j^2 x_r \epsilon_{ri}$ |
| 3 | $d_{jkli}^3 - \frac{2}{3}\delta_{ij}d_{lr}^3\epsilon_{kr} + d_{jk}^3\epsilon_{li} + \alpha^3\delta_{jk}\epsilon_{li}$ | $4(d_{ji}^3 + \alpha^3\delta_{ji})$ | $x_j x_k x_l d_{jkli}^3 - \frac{1}{3}(x_j x_k d_{jr}^3\epsilon_{kr}x_i + R^2 x_j d_{jr}^3\epsilon_{ir}) + (x_j x_k d_{jk}^3 + \alpha^3 R^2)x_l\epsilon_{li}$ |
| 4 | $d_{jklmi}^4 - \frac{15}{16}\delta_{ij}d_{klr}^4 + \frac{5}{24}K_{ijkl}\epsilon_{r\hat{m}}u_r^4$ $+d_{j\hat{k}\hat{l}}^4\epsilon_{\hat{m}i} + K_{j\hat{k}lr}u_r^4\epsilon_{\hat{m}i}$ | $5(d_{jki}^4 + \delta_{j\hat{k}}u_i^4)$ | $x_j x_k x_l x_m d_{jklmi}^4 - \frac{3}{16}x_l x_m(2x_i x_k d_{klr}^4 + 3R^2 d_{ilr}^4)\epsilon_{mr}$ $+\frac{1}{24}(4R^2 x_i x_m\epsilon_{rm} + R^4\epsilon_{ri})u_r^4 + (x_j x_k x_l d_{jkl}^4 + R^2 x_l u_l^4)x_m\epsilon_{mi}$ |

Note: Uses are made of the identity relationship $d_{j\cdots ki}\epsilon_{li} = d_{j\cdots li}\epsilon_{ki}$, $R^2 = x_i x_i$ and $K_{ijkl} = \delta_{ij}\delta_{\hat{k}\hat{l}}$.

Back to the Euler equations (4), we introduce the differential 1-form to write in the form

$$\tilde{p} \triangleq v_k\epsilon_{ki}\Omega dx_i = d\mathcal{H}\tilde{p} + \mathcal{H}d\tilde{p} = -dP. \tag{29}$$

It is obvious to distinguish two parts: one is used to work out the total pressure as

$$P - P_0 = -\mathcal{H}\tilde{p} \tag{30.1}$$

and another can be applied to obtain the further constraints on the tensorial coefficients of the velocity field, say

$$d\tilde{p} = \Omega\partial_j v_k\epsilon_{ki}dx_j \wedge dx_i + v_k\partial_j\Omega\epsilon_{ki}dx_j \wedge dx_i = v_k\partial_k\Omega da = 0 \rightarrow v_k\partial_k\Omega = 0. \tag{30.2}$$

Introducing the general expression of velocity field in terms of finite polynomial, namely

$$v_i(\boldsymbol{x}) = \sum_{K=0}^N v_i^K(\boldsymbol{x}) = \sum_{K=0}^N x_{i_1}\cdots x_{i_K} D_{i_1\cdots i_K i}^K \tag{31.1}$$

with $2 + 3 + 4 + \cdots + (N+2) = \frac{1}{2}(N+1)(N+4)$ independent parameters, and

$$\Omega(\boldsymbol{x}) = \sum_{K=1}^N \Omega^K(\boldsymbol{x}) = \sum_{K=1}^N x_{i_1}\cdots x_{i_{K-1}}\vartheta_{i_1\cdots i_{K-1}}^K, \tag{31.2}$$

we can explicitly work out (30.1) to be

$$P - P_0 = -\mathcal{H}(\tilde{p}) = -\int_0^1 \sum_{K=1}^N \lambda^{K-1} x_{j_1}\cdots x_{j_{K-1}}\vartheta_{j_1\cdots j_{K-1}}^K \sum_{L=0}^N \lambda^L x_{i_1}\cdots x_{i_L} D_{i_1\cdots i_L r}^L \epsilon_{ri}x_i d\lambda$$

$$= -\sum_{K=1}^N\sum_{L=0}^N \frac{1}{K+L} x_{i_1}\cdots x_{i_L} x_{j_1}\cdots x_{j_{K-1}}\vartheta_{j_1\cdots j_{K-1}}^K D_{i_1\cdots i_L r}^L \epsilon_{ri}x_i$$

$$= -\sum_{L=1}^{2N} \frac{1}{L} x_{i_1}\cdots x_{i_L}\sum_{M=0}^L \vartheta_{\hat{i}_1\cdots \hat{i}_M}^{M+1} D_{\hat{i}_{M+1}\cdots i_{L-1}r}^{L-M}\epsilon_{r\hat{i}_L}. \tag{32}$$

Substituting (31) into (30.2) yields no constraint for $N = 0, 1$, and

$$\sum_{L=0}^{2N-2} C_{i_1\cdots i_L}^{N,L} x_{i_1}\cdots x_{i_L} = 0, C_{i_1\cdots i_L}^{N,L} = \sum_{M=0}^L (L-M+1)D_{i_1\cdots \hat{i}_M k}^K \vartheta_{k\hat{i}_{M+1}\cdots \hat{i}_L}^{L-M+2} \tag{33}$$

for $N > 1$. The independence of different polynomial terms $x_1^L x_2^{L-M}$ results in

$$C_{i_1\cdots i_L}^{N,L} = 0, L = 0,1,\cdots, 2N-2. \tag{34}$$

Since $C_{i_1\cdots i_L}^{N,L}$ has $L + 1$ independent parameters, (34) totally includes $1 + 2 + \cdots + 2N - 1 = N(2N-1)$ independent relations, more than the independent parameters of the velocity field when $N > 2$.

For examples, (i) when $N = 0, 1$, both the solutions

$$v_i = v_i^0, v_i = v_i^0 + x_j d_{ji}^1 + \alpha^1 x_j \epsilon_{ji} \quad (34.1)$$

naturally satisfy the Euler equations; (ii) when $N = 2$, the six constraints for nine coefficients of the velocity are

$$v_s^0 \vartheta_s^2 = 3 v_s^0 u_s^2 = 0, D_{js}^1 \vartheta_s^2 = 3 d_{js}^1 u_s^2 + 3 \alpha^1 \epsilon_{js} u_s^2 = 0, j = 1,2; \quad (35.1)$$

$$D_{jks}^2 \vartheta_s^2 = 3 d_{jks}^2 u_s^2 + \frac{9}{8}\left(u_j^2 \epsilon_{ks} + u_k^2 \epsilon_{js}\right) u_s^2 = 0, jk = 11,12,22. \quad (35.2)$$

The first scalar relation means two vectors $v_i^0$ and $u_i^2$ should be perpendicular to each other, the second vector relation is able to be used to work out the scalar $\alpha^1$ as

$$\alpha^1 = -\frac{d_{js}^1 u_s^2 \epsilon_{jr} u_r^2}{u_k^2 u_k^2}. \quad (36)$$

The third tensor relation can be reduced to deviator relation since it is easy to verify $D_{22s}^2 \vartheta_s^2 = -D_{11s}^2 \vartheta_s^2$. Except for these, these constraint relationships are not easy to understand or solve. But it looks like (a) $d_{jki}^2$ and $u_i^2$ are mutually restrictive; (b) the directional feature of $d_{ji}^1$ is also limited by $u_i^2$. Thus, the constrains from the Euler equations seems to be not completely independent, and can be partially solved to make clear the relative module and dependency of directional feature. The most crucial fact is that the quadratic polynomial velocity satisfying the Euler equations seems to exist if we fix partial coefficients or their partial features, say $u_i^2, \alpha^1$ and the size of $v_i^0$ along the direction normal to $u_i^2$. Will the velocity in terms of higher-order polynomials still be like this? Especially when the number of constraint conditions increases compared to the number of independent parameters of velocity field that satisfies the incompressible condition, we need more positive information.

### 3.2 Complex variable formulation and solutions

In order to gain a more thorough observation of the constraints given by the Euler equations, we make use of the more compact formulation in terms complex variable and complex parameters. First, the Euler equations (29) and (30.2) can be written by

$$\tilde{p} = \Omega \text{Im}(\bar{v} dz) = -dP, d\tilde{p} = \text{Im}\left(\bar{v}\frac{\partial \Omega}{\partial \bar{z}} d\bar{z} \wedge dz\right) = 2\text{Re}\left(v\frac{\partial \Omega}{\partial z}\right) da = 0, \quad (37)$$

where use is made of the incompressible condition $\partial_i v_i = 2\text{Re}\left(\frac{\partial v}{\partial z}\right) = 0$. Following (20), we use complex parameters $c_j^K$ to replace the $J$th-order deviator in the $K$th-degree homogeneous polynomial velocity, and get the complex representations of (31) as

$$v = \sum_{K=0}^{N} v^K = \sum_{K=0}^{N}\left(c_{K+1}^K \bar{z}^K + \frac{K+1}{4}\iota \sum_{L=1}^{K}\frac{1}{L} c_{K+1-2L}^K \bar{z}^{K-L} z^L\right), \bar{c}_{-L}^K = c_L^K; \quad (38.1)$$

$$\Omega = 2\text{Im}\left(\frac{\partial v}{\partial z}\right) = \sum_{K=1}^{N} \Omega^K = \sum_{K=1}^{N}\frac{K+1}{2}\sum_{L=1}^{K} c_{K+1-2L}^K \bar{z}^{K-L} z^{L-1}. \quad (38.2)$$

It is easy to verify the incompressible condition

$$\partial_i v_i = 2\text{Re}\left(\frac{\partial v}{\partial z}\right) = 2\text{Re}\left[\iota c_0^1 + \iota \sum_{K=1}^{N}\frac{K+1}{2}\sum_{L=1}^{K} c_{K+1-2L}^K R^{L-1} \bar{z}^{K+1-2L}\right] \equiv 0$$

due to (38.1)$_2$. The total pressure can be integrated from (37)$_1$ as

$$P - P_O = -\mathcal{H}(\tilde{p}) = -\text{Im}\int_0^1 \sum_{J=1}^{N} \lambda^{J-1} \Omega^J \sum_{K=0}^{N} \lambda^K \bar{v}^K z d\lambda = -\text{Im}\sum_{J=1}^{N}\sum_{K=0}^{N}\frac{1}{J+K}\bar{v}^K \Omega^J z. \quad (39)$$

The complex expressions of homogeneous polynomial velocities of degrees up to 4 and their relations to tensorial expressions are listed in Table 2.

Table 2. Complex expressions of homogeneous polynomial velocities $v^K$ satisfying $\text{Re}\frac{\partial v^K}{\partial z} = 0$

| $K$ | $v^K$ | $\Omega^K$ | correspondence |
| --- | --- | --- | --- |
| 1 | $c_2^1 \bar{z} + \frac{1}{2}\iota c_0^1 z$ | $c_0^1$ | $c_2^1 = d_{11}^1 + \iota d_{12}^1, c_0^1 = 2\alpha^1$ |
| 2 | $c_3^2 \bar{z}^2 + \frac{3}{4}\iota\left(c_1^2 z\bar{z} + \frac{1}{2}\bar{c}_1^2 z^2\right)$ | $3\text{Re}(c_1^2 \bar{z})$ | $c_3^2 = d_{111}^2 + \iota d_{112}^2, c_1^2 = u_1^2 + \iota u_2^2$ |
| 3 | $c_4^3 \bar{z}^3 + \iota\left(c_2^3 z\bar{z}^2 + \frac{1}{3}\bar{c}_2^3 z^3 + \frac{1}{2}c_0^3 z^2\bar{z}\right)$ | $4\text{Re}\left(c_2^3 \bar{z}^2 + \frac{1}{2}c_0^3 z\bar{z}\right)$ | $c_4^3 = d_{1111}^3 + \iota d_{1112}^3, c_2^3 = d_{11}^3 + \iota d_{12}^3, c_0^3 = 2\alpha^3$ |
| 4 | $c_5^4 \bar{z}^4 + \frac{5}{4}\iota\left(c_3^4 z\bar{z}^3 + \frac{1}{4}\bar{c}_3^4 z^4\right) + \frac{5}{4}\iota\left(\frac{1}{3}\bar{c}_1^4 z^3\bar{z} + \frac{1}{2}c_1^4 z^2\bar{z}^2\right)$ | $5\text{Re}(c_3^4 \bar{z}^3 + c_1^4 z\bar{z}^2)$ | $c_5^4 = d_{11111}^4 + \iota d_{11112}^4, c_3^4 = d_{111}^4 + \iota d_{112}^4, c_1^4 = u_1^4 + \iota u_2^4$ |

Substituting (38) into (37)$_2$ yields

$$v\frac{\partial \Omega}{\partial z} \equiv 0 \tag{40.1}$$

for $N = 0, 1$, and

$$v\frac{\partial \Omega}{\partial z} = 2\left[\sum_{K=0}^{N}\left(c_{K+1}^{K}\bar{z}^{K} + \frac{K+1}{4}\iota\sum_{L=1}^{K}\frac{1}{L}c_{K+1-2L}^{K}\bar{z}^{K-L}z^{L}\right)\right]\sum_{J=2}^{N}\sum_{M=2}^{J}(M-1)c_{J+1-2M}^{J}\bar{z}^{J-M}z^{M-2} \triangleq \sum_{K=0}^{2N-2}\sum_{L=0}^{K}C_{K-2L}^{N,K}z^{L}\bar{z}^{K-L} \tag{40.2}$$

for $N > 1$. Here

$$C_{K-2L}^{N,K} = \sum_{J=0}^{\min(N,K)}(L+1)c_{J+1}^{J}c_{L-J-1-2K}^{L-J+2} + \iota\sum_{J=0}^{\min(N,K)}\sum_{M=1}^{\min(L,K-J)}\frac{L-M+1}{M}c_{K-J-1-2(L-M)}^{K-J+2}c_{J+1-2M}^{J}. \tag{41.1}$$

The Euler equations (37)$_2$ become

$$C_{K-2L}^{N,K} = \bar{C}_{2L-K}^{N,K}, N > 1, K = 0,1,\cdots,2N-2; L = 0,1,\cdots,K. \tag{41.2}$$

For $N = 4$, we list the constraint relations (41) and their solutions in Table 3. It should be remarked that the solving process must be progressively reduced from the highest-order constraint relations since the highest-order constraint relations are simply involved with the coefficients of the homogeneous part of velocity field. There are three cases to be considered:

Case 1: general solutions with nonzero vorticity where the coefficients in any homogeneous part of velocity are all zero or nonzero;

Case 2: potential flows where only $c_{N+1}^{N}$ could be nonzero except $c_0^1$ indicating the constant vorticity;

Case 3: radially symmetric vorticity where only $c_0^{2M+1}$ could be nonzero.

The obtained general solutions have the following features:
- the modulus of every homogeneous velocity is free, but the magnitude relations of coefficients in the same homogeneous velocity are completely determine;
- the phases of coefficients in the same homogeneous velocity are in a multiple relationship, while the minimum phases of different homogeneous velocities are completely related, say in Table 3:

$$\text{ph}(c_1^2) = \text{ph}(c_1^4), \text{ph}(c_2^3) = \text{ph}(c_2^1) = 2\text{ph}(c_1^4), \text{ph}(c_1^0) = \text{ph}(c_1^4) + \frac{\pi}{2}. \tag{42}$$

Table 3. The constraint relations of quartic polynomial velocity field and the derived solutions

| K | L | bases | $C_{K-2L}^{K} = -\bar{C}_{2L-K}^{K}$ | solutions | | |
|---|---|---|---|---|---|---|
| | | | | Case 1 | Case 2 | Case 3 |
| 0 | 0 | 1 | $\text{Re}(c_1^0\bar{c}_1^2) = 0$ | NS | NS | NS |
| 1 | 0 | $\bar{z}$ | $\frac{3}{2}\bar{c}_1^2c_2^1 + 2c_1^0c_0^3 = \frac{3}{4}\iota c_1^2c_0^1 - 4\bar{c}_1^0c_2^3$ | NS | $c_1^0$ arbitrary | $c_1^0 = 0$ |
| 2 | 0 | $\bar{z}^2$ | $2c_0^3c_1^1 + \frac{3}{2}\bar{c}_1^2c_3^2 + \frac{5}{2}c_1^0c_1^4 = 2\iota c_1^0c_2^2 + \frac{9}{16}\iota c_1^2c_1^2 - \frac{15}{2}\bar{c}_1^0c_3^4$ | $\frac{c_1^0}{\bar{c}_1^0} = -\frac{c_1^2}{\bar{c}_1^2} = -\frac{c_1^4}{\bar{c}_1^4}$ | $c_0^1, c_2^1$ arbitrary | |
| 2 | 1 | $z\bar{z}$ | $\text{Re}\left(4c_2^1\bar{c}_2^3 + \iota c_0^1c_0^3 + \frac{9}{8}\iota c_1^2\bar{c}_1^2 + 5\bar{c}_1^4c_1^0\right) = 0$ | NS | NS | NS |
| 3 | 0 | $\bar{z}^3$ | $2c_0^3c_3^2 + \frac{3}{2}\bar{c}_1^2c_4^3 + \frac{5}{2}c_2^1c_1^4 = 2\iota c_1^2c_2^3 + \frac{15}{4}\iota c_0^1c_3^4$ | $c_2^1 = \iota\frac{c_2^3}{c_0^3}c_0^1 = \iota\frac{c_1^4}{\bar{c}_1^4}\frac{c_0^1}{2}$ | NS | NS |
| 3 | 1 | $z\bar{z}^2$ | $4\bar{c}_2^3c_3^2 + 5c_2^1\bar{c}_1^4 = \frac{3}{2}\iota\bar{c}_1^2c_2^3 - \frac{15}{2}c_3^4\bar{c}_2^1 + \frac{5}{4}\iota c_0^1c_1^4$ | NS | NS | NS |
| 4 | 0 | $\bar{z}^4$ | $\frac{5}{2}c_3^2c_1^4 + 2c_0^3c_4^3 + \frac{3}{2}\bar{c}_1^2c_5^4 = \frac{4}{3}\iota c_2^3c_2^3 + \frac{105}{32}\iota c_1^2c_3^4$ | $\frac{c_1^2}{\bar{c}_1^2} = \frac{c_1^4}{\bar{c}_1^4}, c_3^2 = \frac{3}{8}\iota\frac{c_1^4}{\bar{c}_1^4}c_1^2$ | $c_3^2 \neq 0$ $c_1^2 = 0$ | $c_3^2 = 0$ $c_1^2 = 0$ |
| 4 | 1 | $z\bar{z}^3$ | $5\bar{c}_1^4c_3^2 + 4\bar{c}_2^3c_4^3 = \frac{15}{4}\iota c_3^4\bar{c}_1^2 + \frac{2}{3}\iota c_2^3c_0^2 + \frac{5}{8}\iota c_1^4c_1^2$ | | | |
| 4 | 2 | $z^2\bar{z}^2$ | $\text{Re}\left[\frac{15}{2}\left(c_3^2\bar{c}_3^4 + \frac{1}{2}\iota c_1^2\bar{c}_1^4 + \frac{1}{4}\iota c_1^4\bar{c}_1^2\right) + (4\iota\bar{c}_2^3c_2^3 + \iota c_0^3\bar{c}_0^3)\right] = 0$ | NS | NS | NS |
| 5 | 0 | $\bar{z}^5$ | $2c_0^3c_5^4 + \frac{5}{2}c_1^4c_4^3 = \frac{15}{4}\iota c_3^2c_4^3$ | $c_2^3 = \frac{c_1^4}{\bar{c}_1^4}\frac{c_0^3}{2}, c_4^3 = \frac{\iota}{3}\frac{c_1^4}{\bar{c}_1^4}c_2^3$ | $c_4^3 \neq 0$ $c_0^3 = c_2^3 = 0$ | $c_0^3 \neq 0$ $c_0^3 = c_2^3 = 0$ |
| 5 | 1 | $z\bar{z}^4$ | $4\bar{c}_2^3c_5^2 + 5\bar{c}_1^4c_4^3 = \frac{15}{8}\iota c_3^4c_3^2 + \frac{5}{6}\iota c_2^3c_1^4$ | | | |
| 5 | 2 | $z^2\bar{z}^3$ | $\frac{3}{2}\bar{c}_3^4c_4^3 = \frac{1}{2}\iota c_3^4\bar{c}_2^3 + \frac{1}{6}\iota c_0^3c_1^4 - \frac{1}{3}\iota c_2^3\bar{c}_1^4$ | NS | NS | NS |
| 6 | 0 | $\bar{z}^6$ | $c_1^4c_5^4 = \frac{15}{16}\iota c_3^4c_3^4$ | $c_3^4 = \frac{1}{3}\frac{c_1^4}{\bar{c}_1^4}c_1^4, c_5^4 = \frac{5\iota}{16}\frac{c_1^4}{\bar{c}_1^4}c_1^4$ | $c_5^4 \neq 0$ $c_1^4 = c_3^4 = 0$ | $c_5^4 = 0$ $c_1^4 = c_3^4 = 0$ |
| 6 | 1 | $z\bar{z}^5$ | $\bar{c}_1^4c_5^4 = \frac{5}{16}\iota c_1^4c_3^4$ | | | |
| 2 | $z^2\bar{z}^4$ | $3\bar{c}_3^4 c_5^4 = \frac{5}{4}\iota\left(\frac{1}{6}c_1^4 c_1^4 - \frac{1}{4}c_3^4 \bar{c}_1^4\right)$ | NS | NS | NS |
| 3 | $z^3\bar{z}^3$ | $\text{Re}\left[\frac{25}{8}\iota\left(3c_3^4\bar{c}_3^4 + \frac{4}{3}c_1^4\bar{c}_1^4\right)\right] = 0$ | AS | AS | AS |

Note: "NS" indicates "naturally satisfied", "AS" indicates "always satisfied".

We also list the constraint relations for $N = 2,3$ and their solutions in Table 4 and 5, respectively, and find that the solutions can be read from Table 3 directly though the constraint relations used to get the same solution seem to be different. For example, the results $\frac{c_1^2}{\bar{c}_1^2} = \frac{2c_2^3}{c_0^3}, c_3^2 = \frac{3\iota}{4}\frac{c_2^3}{c_0^3}c_1^2$ come from $C_4^{4,4} = -\bar{C}_{-4}^{4,4}, C_2^{4,4} = -\bar{C}_{-2}^{4,4}$ in Table 3 but from $C_3^{3,3} = -\bar{C}_{-3}^{3,3}, C_1^{3,3} = -\bar{C}_{-1}^{3,3}$ in Table 4.

In order to verify the above derivations, we check (37)$_1$ from the integration of total pressure in (39), and find by substituting the velocity field listing in Table 3 that the necessary condition for $\tilde{p} + dP = 0$ is that $\text{Re}\left(v\frac{\partial\Omega}{\partial z}\right) = 0$ or equivalently (41) holding. On the other hand, we know that the constraint relations (41) construct a nonlinear coupling between the velocity coefficients, which seem to be solvable. The first step to derive the solutions is to work out the highest constraints among the coefficients of highest homogeneous velocity fields. Here let's delve deeper into this.

Table 4. The constraint relations of quadratic polynomial velocity field and the derived solutions

| K | L | bases | $C_{K-2L}^K = -\bar{C}_{2L-K}^K$ | solutions |
|---|---|---|---|---|
| 0 | 0 | 1 | $\text{Re}(c_1^0\bar{c}_1^2) = 0$ | $\bar{c}_1^2 c_1^0 = -c_1^2\bar{c}_1^0$ |
| 1 | 0 | $\bar{z}$ | $\frac{3}{2}\bar{c}_1^2 c_2^1 = \frac{3}{4}\iota c_1^2 c_0^1$ | $c_2^1 = \iota\frac{c_1^2}{\bar{c}_1^2}\frac{c_0^1}{2}$ |
| 2 | 0 | $\bar{z}^2$ | $\frac{3}{2}\bar{c}_1^2 c_3^2 = \frac{9}{16}\iota c_1^2 c_1^2$ | $c_3^2 = \frac{3\iota}{8}\frac{c_1^2}{\bar{c}_1^2}c_1^2$ |
| 2 | 1 | $z\bar{z}$ | $\text{Re}\left(\frac{9}{8}\iota c_1^2\bar{c}_1^2\right) = 0$ | AS |

Note: "AS" indicates "always satisfied".

Table 5. The constraint relations of cubic polynomial velocity field and the derived solutions

| K | L | bases | $C_{K-2L}^K = -\bar{C}_{2L-K}^K$ | solutions |
|---|---|---|---|---|
| 0 | 0 | 1 | $\text{Re}(c_1^0\bar{c}_1^2) = 0$ | NS |
| 1 | 0 | $\bar{z}$ | $\frac{3}{2}\bar{c}_1^2 c_2^1 + 2c_1^0 c_0^3 = \frac{3}{4}\iota c_1^2 c_0^1 - 4\bar{c}_1^0 c_2^3$ | $\frac{c_1^0}{\bar{c}_1^0} = -\frac{c_1^2}{\bar{c}_1^2} = -\bar{c}_1^0\frac{2c_2^3}{c_0^3}$ |
| 2 | 0 | $\bar{z}^2$ | $2c_0^3 c_2^1 + \frac{3}{2}\bar{c}_1^2 c_3^2 = 2\iota c_0^1 c_2^3 + \frac{9}{16}\iota c_1^2 c_1^2$ | $c_2^1 = \iota\frac{c_2^3}{c_0^3}c_0^1$ |
| 2 | 1 | $z\bar{z}$ | $\text{Re}\left(4c_2^1\bar{c}_2^3 + \iota c_0^1 c_0^3 + \frac{9}{8}\iota c_1^2\bar{c}_1^2\right) = 0$ | NS |
| 3 | 0 | $\bar{z}^3$ | $2c_0^3 c_3^2 + \frac{3}{2}\bar{c}_1^2 c_4^3 = 2\iota c_1^2 c_2^3$ | $\frac{c_1^2}{\bar{c}_1^2} = \frac{2c_2^3}{c_0^3}, c_3^2 = \frac{3\iota}{4}\frac{c_2^3}{c_0^3}c_1^2$ |
| 3 | 1 | $z\bar{z}^2$ | $4\bar{c}_2^3 c_3^2 = \frac{3}{2}\iota\bar{c}_1^2 c_2^3$ | |
| 4 | 0 | $\bar{z}^4$ | $2c_0^3 c_4^3 = \frac{4}{3}\iota c_2^3 c_2^3$ | $4c_2^3\bar{c}_2^3 = c_0^3 c_0^3, c_4^3 = \frac{\iota}{6}\frac{c_2^3}{\bar{c}_2^3}c_0^3$ |
| 4 | 1 | $z\bar{z}^3$ | $4\bar{c}_2^3 c_4^3 = \frac{2}{3}\iota c_2^3 c_0^3$ | |
| 4 | 2 | $z^2\bar{z}^2$ | $\text{Re}(4\iota\bar{c}_2^3 c_2^3 + \iota c_0^3 c_0^3) = 0$ | AS |

Note: "NS" indicates "naturally satisfied", "AS" indicates "always satisfied".

Replacing $c_{N+1}^N$ by $\frac{\iota(N+1)}{4}c_{N+1}^N$, (38.1) with only the highest degree becomes

$$v = v^N = \frac{\iota(N+1)}{4}\left(c_{N+1}^N\bar{z}^N + \sum_{L=1}^N\frac{1}{L}c_{N+1-2L}^N\bar{z}^{N-L}z^L\right), \bar{c}_{-L}^N = c_L^N; \quad (43.1)$$

while

$$\Omega = \Omega^N = 2\text{Im}\left(\frac{\partial v^N}{\partial z}\right) = \frac{N+1}{2}\sum_{L=1}^N c_{N+1-2L}^N\bar{z}^{N-L}z^{L-1}. \quad (43.2)$$

After some derivation and organization, we get the constraints from $\mathrm{Re}\left(v^N \frac{\partial \Omega^N}{\partial z}\right) = 0$ as

$$(L+1)c_{N+1}^N c_{N-3-2L}^N + \sum_{K=1}^{L} \frac{L-K+1}{K} c_{N+1-2K}^N c_{N-3-2(L-K)}^N = \sum_{K=0}^{\min(N-L,L)} \frac{N+K-L-1}{N-K} c_{N-2K-1}^N c_{N-1-2(L-K)}^N \quad (44)$$

for $L = 0, \cdots, N-2$. According to the practice before, we may further assume that $c_K^N = a_K^N e^{\iota K \varphi}, a_{-K}^N = a_K^N \in \mathbb{R}$, and obtain

$$(L+1)a_{N+1}^N a_{N-2L-3}^N = \frac{N-1}{N-L} a_{N-1}^N a_{N-2L-1}^N + \sum_{K=0}^{L-1} \left( \frac{N+K-L-1}{N-K} - \frac{L-K}{K+1} \right) a_{N-2K-1}^N a_{N-1-2(L-K)}^N \quad (45)$$

for $L = 0, \cdots, \left[\frac{N+1}{2}\right] - 1$. We list the constraint relations and the corresponding solutions in Table 6 to demonstrate the solvability of (45), where there is a free scalar for all $N$.

Table 6. The constraint relations of homogeneous polynomial velocities and the derived solutions

| $N$ | $L$ | constraints | solutions |
|---|---|---|---|
| 2 | 0 | $a_3^2 a_{-1}^2 = \frac{1}{2} a_1^2 a_1^2$ | $a_3^2 = \frac{1}{2} a_1^2$ |
| 3 | 0 | $a_4^3 a_0^3 = \frac{2}{3} a_2^3 a_2^3$ | $a_2^3 a_2^3 = \frac{1}{4} a_0^3 a_0^3, a_4^3 = \frac{1}{6} a_0^3$ |
| 3 | 1 | $2 a_4^3 a_{-2}^3 = \frac{1}{3} a_2^3 a_0^3$ | |
| 4 | 0 | $a_5^4 a_1^4 = \frac{3}{4} a_3^4 a_3^4$ | $a_5^4 = \frac{1}{4} a_3^4, a_3^4 = \frac{1}{3} a_1^4$ |
| 4 | 1 | $2 a_5^4 a_{-1}^4 = \frac{5}{6} a_3^4 a_1^4$ | |
| 5 | 0 | $a_6^5 a_2^5 = \frac{4}{5} a_4^5 a_4^5$ | $a_2^5 a_2^5 = \frac{4}{9} a_0^5 a_0^5, a_4^5 = \frac{1}{6} a_0^5, a_6^5 = \frac{1}{20} a_2^5$ |
| 5 | 1 | $2 a_6^5 a_0^5 = \frac{3}{5} a_4^5 a_2^5$ | |
| 5 | 2 | $3 a_6^5 a_{-2}^5 = \frac{1}{4} a_2^5 a_2^5 - \frac{4}{15} a_4^5 a_0^5$ | |
| 6 | 0 | $a_7^6 a_3^6 = \frac{5}{6} a_5^6 a_5^6$ | $a_3^6 = \frac{1}{2} a_1^6, a_5^6 = \frac{1}{10} a_1^6, a_7^6 = \frac{1}{60} a_1^6$ |
| 6 | 1 | $2 a_7^6 a_1^6 = \frac{2}{3} a_5^6 a_3^6$ | |
| 6 | 2 | $3 a_7^6 a_{-1}^6 = \frac{3}{10} a_3^6 a_3^6 - \frac{1}{4} a_5^6 a_1^6$ | |

In summary, besides the constant and linear velocity fields satisfying the Euler equations constantly, we present the following theorem for the general cases of local polynomial velocity fields:

**Theorem**: In two-dimensional spaces, the Euler equations of ideal fluids have analytical solutions in terms of finite polynomials (38.1) with maximal degree $N > 1$, where the complex coefficients $\{c_{K+1-2L}^K, K = 0, 1, \cdots, N; L = 0, 1, \cdots, K\}$ satisfy $c_{-L}^K = \bar{c}_L^K$, and the constraint relations (41). These constraints are speculated to be always solvable, with the form $c_L^K = a_L^K \exp(\iota L \varphi), a_L^K \in \mathbb{R}$. The moduli (scalars) set $\{a_{-L}^K = a_L^K, L = K+1, K-1, \cdots\}$ is free as a whole, but there are fixed mutual constraints within it, which can be deconstructed beginning from the highest nonlinear self-coupling constraints. Therefore, there are totally $N + 2$ independent parameters, including $N + 1$ modulus parameters and one phase parameter, for $N$th-degree polynomial velocity field.

## 4 Examples and Discussions

### 4.1 Special case of potential flows

Two-dimensional $N$th-degree polynomial velocity fields have maximal $2 + 4 + \cdots + 2(N+1) = (N+1)(N+2)$ parameters, which reduce to $2 + 3 + \cdots + N + 2 = \frac{1}{2}(N+1)(N+4)$ when the incompressible condition is applied. For potential flows without vorticity, the nonlinear couplings between polynomial velocity fields with different degrees vanish, the Euler equations become the Bernoulli's theorem

$$dP = d\left(p + \frac{1}{2} v_i v_i\right) = 0. \quad (46)$$

The number of independent parameters of polynomial velocity field become $2(N+1)$, from (43.1) that is

$$v = \sum_{K=0}^{N} v^K = \tfrac{1}{2}\iota c_0^1 z + \sum_{K=0}^{N} \tfrac{\iota(N+1)}{4} c_{K+1}^K \bar{z}^K. \tag{47}$$

It is interesting that the number of independent parameters in (47) is larger than the number of independent parameters of polynomial velocity field (38.1) with nonzero vorticity.

**4.2 Special case of radially symmetric vorticity**

We notice a special case where only the coefficients $c_0^{2M+1} \neq 0$, or the vorticity is a function of $z\bar{z} = r^2$. From

$$v = \sum_{K=0}^{M} v^{2K+1} = \frac{\iota}{2} \sum_{K=0}^{M} c_0^{2K+1} \bar{z}^K z^{K+1}, \tag{48}$$

it is easy to prove

$$\mathrm{Re}\left(v \frac{\partial \Omega}{\partial z}\right) = \mathrm{Re}\left(\iota z \sum_{K=0}^{M} \frac{1}{K+1} c_0^{2K+1} r^{2K} \bar{z} \sum_{L=1}^{M} L c_0^{2L+1} r^{2L-2}\right) \equiv 0.$$

More general,

$$\Omega = f(r^2), \text{or } = \iota z F(r^2) \text{ with } x F(x) = \int f(x) dx \tag{49}$$

always satisfies the Euler equations (Majda and Bertozzi, 2002). A well-known example of (49) is the point with infinite velocity but finite circulation, say

$$v = \frac{\iota \Gamma_0}{2\pi} \frac{z}{r^2}, z \in \mathcal{D}\setminus\{0\} \rightarrow \Omega(\boldsymbol{x}) \equiv 0, \Gamma = \mathrm{Re} \oint_{\partial \mathcal{D}} v d\bar{z} \equiv \Gamma_0, \tag{50}$$

where the streamlines are circles centered at the origin, called the center-type isotropic point. However, in this situation the velocity field is no longer expressible in finite polynomials.

**4.2 the N-S equations**

For polynomial velocity fields, it is easy to extend to the N-S equations. Using the third-degree polynomial velocity field as an example, from

$$v = \sum_p v^p = \tfrac{1}{4}\iota c_1^0 + \tfrac{1}{2}\iota c_2^1 \bar{z} + \tfrac{1}{2}\iota c_0^1 z + \tfrac{3}{4}\iota c_3^2 \bar{z}^2 + \tfrac{3}{4}\iota c_1^2 z\bar{z} + \tfrac{3}{8}\iota \bar{c}_1^2 z^2 + \iota c_4^3 \bar{z}^3 + \iota c_2^3 z\bar{z}^2 + \tfrac{1}{2}\iota c_0^3 z^2 \bar{z} + \tfrac{1}{3}\iota \bar{c}_2^3 z^3 \tag{51.1}$$

we have

$$\Omega = 2\mathrm{Im}\left(\frac{\partial v}{\partial z}\right) = \sum_p \Omega^p = c_0^1 + \tfrac{3}{2}(\bar{c}_1^2 z + c_1^2 \bar{z}) + 2(\bar{c}_2^3 z^2 + c_0^3 z\bar{z} + c_2^3 \bar{z}^2) \tag{51.2}$$

and

$$\nabla^2 v = 4\frac{\partial^2 v}{\partial z \partial \bar{z}} = 3\iota c_1^2 + 8\iota(c_2^3 \bar{z} + c_0^3 z), \nabla^2 \Omega = 8c_0^3. \tag{51.3}$$

The complex form

$$\begin{cases} 2\mathrm{Re}\left(\frac{\partial v}{\partial z}\right) = 0, \\ \Omega \mathrm{Im}(\bar{v}dz) - \nu \mathrm{Re}(\nabla^2 \bar{v}dz) = -dP \rightarrow \mathrm{Re}\left(v\frac{\partial \Omega}{\partial z}\right) + \nu \nabla^2 \Omega = 0. \end{cases} \tag{52.1}$$

of the steady N-S equations

$$\begin{cases} \partial_k v_k = 0, \\ v_k \partial_k v_i = -\partial_i p + \nu \nabla^2 v_i. \end{cases} \tag{52.2}$$

yields the same constraint relations in Table 4 except the constant term becoming

$$\mathrm{Re}(c_1^0 \bar{c}_1^2) = 0 \Rightarrow \mathrm{Re}(c_1^0 \bar{c}_1^2 + 8\nu c_0^3) = 0. \tag{53}$$

As pointed out in the previous section, this relation is not isolated, such a requirement must destroy the derived solutions, or simply say, (53) will result in no solution of finite polynomials. So, the N-S equations is probably incompatible with finite polynomial velocity fields.

**4.3 Significance of finite polynomial solutions**

There are no ideal fluids in real word, but both water and gas have weak viscosity and weak compressibility. That means in steady flows far away from the boundaries, the solutions of the Euler equations present almost the real state of flow. In addition, the solutions of the Euler equations centrally reflect the role of nonlinear convective term, which is an issue that cannot be avoided in solving any other flow theories. Finally, when the flow geometry is complicated, it is impossible to obtain the analytical solutions and the numerical treatment seems imperative. At this point, the global solution is decomposed into lots of local solutions, which can be expressed by the finite polynomial solutions obtained in this paper.

**4.4 Examples and discussions**

According to the results obtained in the previous section, we may assume

$$c_1^{2M} = \frac{A_{2M}}{M-1} e^{\iota\varphi}, c_0^{2M+1} = \frac{A_{2M+1}}{M}, c_2^{2M+1} = \frac{M}{M+1} c_0^{2M+1} e^{\iota 2\varphi} = \frac{A_{2M+1}}{M+1} e^{\iota 2\varphi} \tag{54}$$

to write the quartic polynomial velocity field in the form

$$v = \tfrac{1}{4}\iota A_0 e^{\iota\varphi} + \tfrac{1}{2}\iota A_1 e^{\iota\varphi}(\bar{z}e^{\iota\varphi} + ze^{-\iota\varphi}) + \tfrac{3}{4}\iota A_2 e^{\iota\varphi}\left(\tfrac{1}{2}\bar{z}^2 e^{\iota 2\varphi} + z\bar{z} + \tfrac{1}{2}z^2 e^{-\iota 2\varphi}\right)$$
$$+ \iota A_3 e^{\iota\varphi}\left(\tfrac{1}{6}\bar{z}^3 e^{\iota 3\varphi} + \tfrac{1}{2}z\bar{z}^2 e^{\iota\varphi} + \tfrac{1}{2}z^2\bar{z}e^{-\iota\varphi} + \tfrac{1}{6}z^3 e^{-\iota 3\varphi}\right)$$
$$+ \tfrac{5}{4}\iota A_4 e^{\iota\varphi}\left(\tfrac{1}{12}\bar{z}^4 e^{\iota 4\varphi} + \tfrac{1}{3}z\bar{z}^3 e^{\iota 2\varphi} + \tfrac{1}{2}z^2\bar{z}^2 + \tfrac{1}{3}z^3\bar{z}e^{-\iota 2\varphi} + \tfrac{1}{12}z^4 e^{-\iota 4\varphi}\right). \tag{55}$$

Then, from

$$\Omega = 2\mathrm{Im}\left(\frac{\partial v}{\partial z}\right) = A_1 + \tfrac{3}{2}A_2(\bar{z}e^{\iota\varphi} + ze^{-\iota\varphi}) + 2A_3\left(\tfrac{1}{2}\bar{z}^2 e^{\iota 2\varphi} + z\bar{z} + \tfrac{1}{2}z^2 e^{-\iota 2\varphi}\right)$$
$$+ \tfrac{5}{2}A_4\left(\tfrac{1}{3}\bar{z}^3 e^{\iota 3\varphi} + z\bar{z}^2 e^{\iota\varphi} + z^2\bar{z}e^{-\iota\varphi} + \tfrac{1}{3}z^3 e^{-\iota 3\varphi}\right), \tag{56.1}$$

$$\frac{\partial\Omega}{\partial z} = \tfrac{3}{2}A_2 e^{-\iota\varphi} + 2A_3 e^{-\iota\varphi}(\bar{z}e^{\iota\varphi} + ze^{-\iota\varphi}) + \tfrac{5}{2}A_4 e^{-\iota\varphi}(\bar{z}^2 e^{\iota 2\varphi} + 2z\bar{z} + z^2 e^{-\iota 2\varphi}), \tag{56.2}$$

it is easy to prove

$$\mathrm{Re}\left(v\frac{\partial\Omega}{\partial z}\right) = \mathrm{Re}\left\{\iota\left[\tfrac{1}{4}A_0 + \tfrac{1}{2}A_1(\bar{z}e^{\iota\varphi} + ze^{-\iota\varphi}) + \tfrac{3}{4}A_2\left(\tfrac{1}{2}\bar{z}^2 e^{\iota 2\varphi} + z\bar{z} + \tfrac{1}{2}z^2 e^{-\iota 2\varphi}\right)\right.\right.$$
$$+ A_3\left(\tfrac{1}{6}\bar{z}^3 e^{\iota 3\varphi} + \tfrac{1}{2}z\bar{z}^2 e^{\iota\varphi} + \tfrac{1}{2}z^2\bar{z}e^{-\iota\varphi} + \tfrac{1}{6}z^3 e^{-\iota 3\varphi}\right)$$
$$+ \tfrac{5}{4}A_4\left(\tfrac{1}{12}\bar{z}^4 e^{\iota 4\varphi} + \tfrac{1}{3}z\bar{z}^3 e^{\iota 2\varphi} + \tfrac{1}{2}z^2\bar{z}^2 + \tfrac{1}{3}z^3\bar{z}e^{-\iota 2\varphi} + \tfrac{1}{12}z^4 e^{-\iota 4\varphi}\right)\right]\left[\tfrac{3}{2}A_2 + 2A_3(\bar{z}e^{\iota\varphi} + ze^{-\iota\varphi})\right.$$
$$\left.\left.+ \tfrac{5}{2}A_4(\bar{z}^2 e^{\iota 2\varphi} + 2z\bar{z} + z^2 e^{-\iota 2\varphi})\right]\right\} = \mathrm{Re}(\iota \times \mathrm{real} \times \mathrm{real}) \equiv 0.$$

The $N+2$ system $\{A_0, A_1, A_2, A_3, A_4; \varphi\}$, with five scalars and an angular parameter, establishes a local polynomial velocity field that satisfies the Euler equations everywhere, but reveal a shocking fact that the velocity must have a unitary direction in the region. Further, if we substitute $z = re^{\iota\theta}, n = e^{\iota\varphi}$ into (52), the expression

$$v = \tfrac{1}{4}\iota A_0 e^{\iota\varphi} + \iota A_1 e^{\iota\varphi} r\cos(\theta - \varphi) + \tfrac{3}{2}\iota A_2 e^{\iota\varphi} r^2 \cos^2(\theta - \varphi) + \tfrac{4}{3}\iota A_3 e^{\iota\varphi} r^3 \cos^3(\theta - \varphi) + \tfrac{5}{3}\iota A_4 e^{\iota\varphi} r^4 \cos^4(\theta - \varphi) \tag{57}$$

indicates that the velocity simply depends on the distance along the direction perpendicular to the flow, namely the flow is a simple shear flow.

As well-known, for the steady flows of ideal fluids, the vorticity keeps invariant along the streamline, that means the gradient of it must be normal to the streamline while the local straight streamlines come from the nonlinear constraint of convective term. For the streamlines have no vorticity at their starting points, the region they occupy must be potential flow. The velocity field (47) of potential flow consists of harmonic functions in polynomial form. According to the fundamental theorem of algebra of (47), there must be some zero points of velocity, without loss of generality, let the origin be a zero point, $c_1^0 = 0$. Then, the origin becomes an isotropic point and its local topological property of streamlines is determined by the lowest nonzero coefficient in (47), say $v \sim c_{K+1}^K \bar{z}^K$. Using $z = re^{\iota\theta}, c_{K+1}^K = ce^{\iota\phi}, c \in \mathbb{R}^+, 0 \leq \theta, \phi < 2\pi$, we know there are always some streamlines into [out of] the origin, satisfying

$$\phi - K\theta_I = \theta_I - I\pi, \qquad I = 0, 1, \cdots, 2K - 1. \tag{58}$$

The origin is a saddle in this case.

Reninding the general method presented by Majda and Bertozzi (2002) to construct exact steady solutions of the 2D Euler Equations, that is the simple dependence of vorticity on the stream function $\psi(z,\bar{z})$. Assume $v = 2\iota \frac{\partial \psi}{\partial \bar{z}}$, we have

$$\Omega = 2\text{Im}\left(\frac{\partial v}{\partial z}\right) = \nabla^2 \psi = \Omega(\psi). \tag{59}$$

The study in this paper confirms that if the stream function is assumed to be a polynomial, there are three types of solutions:

$$\psi = (x_i n_i)^K, (x_i x_i)^K, harmonic\ polynomials, \tag{60}$$

where the local distributions of vorticity are schistose, ring-shaped, and blocky (constant or zero), respectively.

## 5 Conclusions

In this paper, the local finite polynomial solutions of the two-dimensional steady Euler equations are studied. Due to the invariance of vorticity along the streamline, the flow domain can be clearly divided into starshaped regions with or without vorticity. In virtue of the tensorial representation of velocity coefficients and complex variable expression, the $N$th-degree polynomial velocity with nonzero vorticity is derived to be determined by $N + 1$ scalars and one angular parameter while the $N$th-degree polynomial velocity without vorticity is determined by $N + 1$ complex parameters. The former is further verified to be a straight shear flow, while the latter may have some saddle-type isotropic points. These results provide vivid examples to understand the nonlinear coupling mechanism in the convective term.

**Declaration of interests**. The author reports no conflict of interest.